\documentclass{aa} 
\usepackage{natbib}
\bibpunct{(}{)}{;}{a}{}{,} 

\usepackage{graphicx}
\usepackage[normalem]{ulem}   

\usepackage{txfonts}
\begin{document}
   \title{Circumstellar dust shells around long-period variables}
   \subtitle{X. Dynamics of envelopes around standard luminous, C-rich AGB stars}
   \author{ C. Dreyer\inst{}, M. Hegmann\inst{}, \and E. Sedlmayr\inst{}}
   \authorrunning{C. Dreyer et al.}
   \titlerunning{Circumstellar dust shells around long-period variables X.}

   \offprints{C. Dreyer}
   \institute{Technische Universit\"at Berlin, Zentrum f\"ur Astronomie und Astrophysik (ZAA), EW 8-1,
              Hardenbergstr. 36, D-10623 Berlin\\
              \email{dreyer@astro.physik.tu-berlin.de}
              }
   \date{Received 27 August / Accepted 15 October}

  \abstract
  {Long-period variables (LPVs) and Miras exhibit a pronounced variability in their luminosity with a more or 
  less well-defined period, and they suffer large mass loss in the form of stellar winds. Owing to this extensive 
  mass loss, they are surrounded by extended circumstellar dust shells (CDSs). The dynamics of these envelopes 
  is the result of a complex interplay via an external excitation by the pulsating central star, dust formation, 
  and radiative transfer.}
  %
  {Our study is aimed at an understanding of the dynamics of CDSs around carbon-rich, standard luminous LPVs 
  and Miras. These shells often show multiperiodicity with secondary periods as high as a few $10^4\,\mathrm{d}$ 
  superimposed on a main period that is in the range of approximately $10^2-10^3\,\mathrm{d}$. Such secondary 
  periods may be caused at least in part by the presence of dust.}
  %
  {We consider an excitation of the CDSs either by a harmonic force, provided by the oscillation of the central 
  star, or by a stochastic force with a continuous power spectrum. The resulting numerically computed dynamical 
  behaviour of the shell is analysed with the help of Fourier analysis and stroboscopic maps.}
  %
  {CDSs may be described as multioscillatory systems that are driven by the pulsating stars. A set of normal 
  modes can be identified. The obtained periods of these modes are some $10^3\,\mathrm{d}$, which 
  is a characteristic timescale for dust nucleation, growth, and elemental enrichment in the dust formation zone. 
  Depending on the oscillation period and strength of the central star, the envelope reacts periodically, multi-
  periodically, or irregularly.}
  %
  {}
  \maketitle
%
%
%
\section{Introduction}
Long-period variables (LPVs) and Miras are highly evolved stars located in the upper region of the 
asymptotic giant branch (AGB). They are characterised by low surface temperatures $T<3\times 10^3\,
\mathrm{K}$ and high luminosities of the order of $10^4\,\mathrm{L_\odot}$. They typically exhibit 
a pronounced variability of their intensity and suffer large mass loss in the form of a stellar wind 
with mass loss rates of up to $10^{-4}\ \mathrm{M_\odot\,yr^{-1}}$. Owing to their extensive mass loss, 
LPVs and Miras are surrounded by extended circumstellar envelopes that provide the necessary 
conditions for the formation of dust.

\cite{dreyer09} (henceforth called paper I), a previous paper of this series, studied the dynamics of 
circumstellar dust shells (CDSs) around high luminous LPVs and Miras. It turned out that for these kind 
of stars, the dynamics of the envelopes is clearly dominated by the interplay between dust formation and 
radiative transfer, i.e. the exterior $\kappa$-effect \citep[see][]{fleischer95,hoefner95}. A regular 
oscillation of the shell can be solely explained by the formation of dust and the radiative pressure 
on it.

In this work we concentrate on the circumstellar envelopes of less luminous LPVs and Miras. For luminosities 
of less than $10^4\,\mathrm{L_\odot}$, additional input of energy and momentum is required to sustain the 
observed oscillations. This energy is provided by oscillations of the star's deeper layers. Caused by the 
pulsational instability due to variations in the convective flux \citep[cf.][]{fox82}, hydrodynamic waves 
are generated in the dense stellar atmosphere and steepen into shock waves when travelling into the thinner 
circumstellar medium. The presence of such shock waves causes a levitation of the atmosphere producing 
favourable conditions for dust nucleation and growth. Once the first dust particles have formed, the momentum 
coupling between matter and radiation field is strongly enhanced, which again results in a massive stellar wind.

Although the shell's oscillation has to be triggered by a stellar pulsation, the presence of dust influences 
the dynamics of the entire system. Even for a strictly harmonic excitation, the shell does not necessarily 
mimic the stellar pulsation, but instead exhibits its own normal modes \citep[cf.][]{barthes94}. 
Superimposed upon the excitation period that is in the range of approximately $(10^2-10^3)\,\mathrm{d}$ 
\citep{whitelock91}, the envelope typically shows oscillations with periods of up to a few $10^4\,\mathrm{d}$. 
This could explain, at least in part, the observed cycle-to-cycle variation in the periods of some Miras 
\citep{eddington29}. As pointed out by \cite{percy99}, these fluctuations could come from multiperiodicity, 
with either a secondary period that is comparable to the primary one or a secondary period an order of magnitude 
longer than the primary one. In the latter case, the secondary oscillation may be explained by an oscillation of 
the circumstellar shell from the exterior $\kappa$-effect.

In the context of dynamical systems, the CDS can be described as a multioscillatory system that is driven by the 
pulsating stellar atmosphere. The resulting dynamical behaviour depends on both the frequency and strength of the 
excitation, as well as, on the intrinsic dynamics of the shell. For a close examination of the possible normal modes 
of a CDS, which are controlled by the intrinsic timescales of the various physical and chemical processes, we 
consider response spectra of a circumstellar shell that is excited not only by a harmonic force but also by a 
stochastic force with a continuous power spectrum. The inherently non-linear nature of the underlying equations 
introduces an element of randomness into the shell's dynamics. This also becomes clear when the CDS is excited 
by a strictly harmonic force. For a sensitive range of frequencies and excitation strengths, the shell does not 
behave strictly periodically but shows signs of deterministic chaos.

This paper is organised as follows. Section \ref{modelling} outlines our dynamical modelling method of a CDS, 
giving a special focus to the formulation of a purely harmonic and stochastic excitation of the CDS 
provided by a central star. Section \ref{diagnostic} presents our direct methods as used to investigate the dynamical 
CDS behaviour. This section discusses the problems of creating and analysing discrete power spectra and deterministic 
maps. In Sect.~\ref{application}, we apply the method to a reference model. The eigendynamics are first 
determined and then the complex interaction between the inner excitation by a pulsating star and the multioscillatory 
circumstellar shell is discussed, in this case depending on excitation frequency and strength. The paper finishes 
with some concluding remarks in Sect.~\ref{summary}.\\
%
%
%
\section{Modelling of CDSs}\label{modelling}
The theoretical model of a CDS is conceived as the complete unique set of solutions of an appropriate 
system of fundamental dynamical equations, comprising the conservation laws, transport equations, and 
constitutive equations that describe both the local and global physical situation of the CDS and the 
various inherent processes, such as radiative transfer \citep{lucy71,lucy76,unnokondo76} 
or dust formation \citep{gail88,gauger90}. These equations were listed and solved according 
to the procedure given in \cite{fleischer92}. 
The CDSs are characterised by the fundamental parameters of their generating stars, which are mass $M_*$, 
luminosity $L_*$, effective temperature $T_*$, and carbon-to-oxygen abundance ratio C/O. As these parameters 
change only slightly with time because of stellar evolution, they are considered as given external parameters 
for modelling a circumstellar shell.\\
\subsection{Monoperiodic excitation}\label{monoperiodic}
   The pulsation of an AGB star, more specifically the underlying inner $\kappa$-mechanism, is treated 
   by the piston approximation adopted from the work of \cite{wood79}, \cite{bowen88a}, and \cite{bowen91}. 
   It is assumed that the radius of the inner boundary $R_{\mathrm{in}}(t)$ located a few scale heights 
   below the stellar radius oscillates sinusoidally about an equilibrium position $R_0$ with
   \begin{equation}\label{pulsation_r}
      R_{\mathrm{in}}(t)=R_0+\Delta u\frac{P}{2\pi} \sin\left(\frac{2\pi}{P}t\right)\qquad .
   \end{equation}
   This implies that, at the inner boundary the hydrodynamic velocity $u_{\mathrm{in}}$, varies in a manner 
   determined by
   \begin{equation}\label{pulsation_v}
     u_{\mathrm{in}}(t)=\Delta u \cos\left(\frac{2\pi}{P}t\right)\qquad .
   \end{equation}
   The velocity amplitude $\Delta u$, which simulates the strength of the interior pulsation of the star, 
   as well as the pulsation period $P$, are input parameters.
 \subsection{Stochastic excitation}\label{stochastic}
   To examine the eigenfrequencies and the band-pass characteristics of circumstellar 
   shells around standard and low-luminous AGB stars, we consider the excitation of the 
   shell by external white noise as generated by, e.g., a convective atmosphere of the 
   giants. In analogy to the piston approximation, the movement of the inner boundary $R_{\mathrm{in}}(t)$ 
   is described by the following stochastic equations:
   \begin{equation}\label{wiener_r}
     R_{\mathrm{in}}(t)=R_0 + \int_{0}^{t}\Gamma(t')\ {\rm d} t'
   \end{equation}
   \begin{equation}\label{wiener_u}
     u_{\mathrm{in}}(t)=\Gamma(t)\qquad .
   \end{equation}
   The term $\Gamma(t)$ describes the influence of the randomness or noise and is taken 
   to be a Gaussian with a mean of zero:
   \begin{equation}
     \left< \Gamma \right>=0\, .
   \end{equation}
   Since we adopt white noise, $\Gamma(t)$ is uncorrelated in time, which corresponds to 
   the (auto)correlation function $K(\tau)$ being Dirac's $\delta$-function
   \begin{equation}
     K(\tau)=\left< \Gamma(t)\Gamma(t+\tau) \right>=\sigma^2\delta(\tau)\, ,
   \end{equation}
   where $\sigma^2$ scales the intensity of the noise. According to the Wiener-Khinchin 
   theorem, the spectral density $|H(f)|^2$ of $\Gamma(t)$ is given by the Fourier transform 
   of its autocorrelation function,
   \begin{equation}
      |H(f)|^2=\int_{-\infty}^{+\infty} K(\tau)\, {\rm e}^{2\pi if\tau}\, {\rm d}\tau=\sigma^2 ,
   \end{equation}
   and is therefore independent of the frequency $f$.
   The integral of the Gaussian white noise, i.e. the solution of
   Eq.~(\ref{wiener_u}), is known as the Wiener process. It is also Gaussian 
   distributed, so the mean vanishes, and its variance increases linearly with time $t$. 
   For a concrete numerical modelling of the CDS, only realisations of the stochastic 
   process Eq.~(\ref{wiener_r}) where $R_{\mathrm{in}}(t)$ stays within predefined finite 
   boundaries (to ensure that the inner boundary stays within the stellar radius $R_*$) are 
   considered. The strength of the noise $\sigma$ is also an input parameter.\\
%
%
%
\section{Diagnostic approach}\label{diagnostic}
The dynamics of (non-)linear systems may be described by means of different characteristics. 
One of the classical ways to characterise the dynamics is to compute the power spectrum based 
on the Fourier transform. This provides the possibility of obtaining a deeper physical 
understanding of the real system dynamics by detecting various frequencies contained 
in a given signal. Another way is an examination by means of a deterministic approach, e.g., to 
create stroboscopic or specific Poincar\'e maps to find domains of stability in the phase space.
\subsection{Fourier transform}
 Generally, the Fourier transform is simply a method of expressing a function in terms of the sum of its projections 
 onto a set of basis functions. The standard basis functions used for Fourier transform are $\{\sin(2 \pi f t)$, 
 $\cos(2 \pi f t), f \in {\mathbb R} \}$ or, equivalently, $\{e^{2 \pi ift}, f \in {\mathbb R} \}$. It is the 
 oscillatory frequency $f$ that varies over the set of all real numbers to give an infinite collection 
 of basis functions.  We project a given time function $h(t)$ onto our basis functions to get the Fourier 
 amplitudes $H(f)$ for each frequency $f$:
 \begin{equation} 
   {\cal F}(h(t)) = H(f) = \int h(t)\, e^{2 \pi i f t} dt. 
 \end{equation}
 In general, $H(f)$ will be complex. The norm of the amplitude, $|H(f)|$ is called the Fourier 
 spectrum, and the square $|H(f)|^2$ is called the power spectrum of $h(t)$.

 When using a fast Fourier-transform (FFT)-algorithm to compute the spectrum of a signal, we 
 use a finite length of signal, consisting of $N=2^{\gamma}, \gamma \in \mathbb{N}$ 
 discrete samples. However, the discretisation of a signal is not  without problems.
 If the sampling rate $f_s=1/\Delta t$ is not high enough to sample the signal correctly, then a 
 phenomenon called aliasing occurs; i.e., components of the signal at high frequencies 
 are mistaken for components at lower frequencies.
\subsection{Mapping}
 For dynamic systems having several degrees of freedom, it is not very practical to discuss the orbit 
 in a multidimensional phase space. A more appropriate way is to study the intersections of orbit 
 with a plane in phase space, which reduces the phase space dimension by one.

 If the differential equations describing a system contain terms periodic in time (where $P$ is 
 the period), it is convenient to create stroboscopic maps showing the time trajectory 
 for several time intervals $\Delta t$; i.e., the flow is sampled whenever $t = n\Delta t$ 
 for $n \in \mathbb{N}_0$. This kind of map can be constructed for any temporal interval.
 Sampling by a fixed time interval $\Delta t=P$, i.e. by successive times when the orbit crosses 
 the phase angle $\phi$ plane, i.e. $t=k2\pi$ with $k$ being an integer, results in a Poincar\'e map.
%
%
%
\section{Application}\label{application}
 The condition for a purely dust-driven wind by the exterior $\kappa$-mechanism is only properly fulfilled 
 at the very end of the AGB evolution. If the stellar luminosity decreases, the dominance of the dust with 
 respect to the CDS dynamics at first diminishes. Then, further reduction eventually leads to a vanishing 
 outflow, if no additional energy input is provided. AGBs with low and standard luminosities generate neither
 self-induced shocks and outflows, with self-maintained oscillatory patterns caused solely by the dust formation 
 (exterior $\kappa$-mechanism), nor a CDS. Generating a CDS requires input of momentum and energy via an 
 additional mechanism in order to levitate material into the outer atmosphere, i.e. into the dust formation 
 window. An example of such a mechanism can be the pulsation of the outer layers of a red giant. The pulsation 
 provides the star with mechanical momentum for levitating the material in their atmosphere.
 \subsection{Eigendynamics of envelopes}\label{eigendynamics_modella14}
\begin{figure}[htb]
    \centering
     \includegraphics[width=9cm]{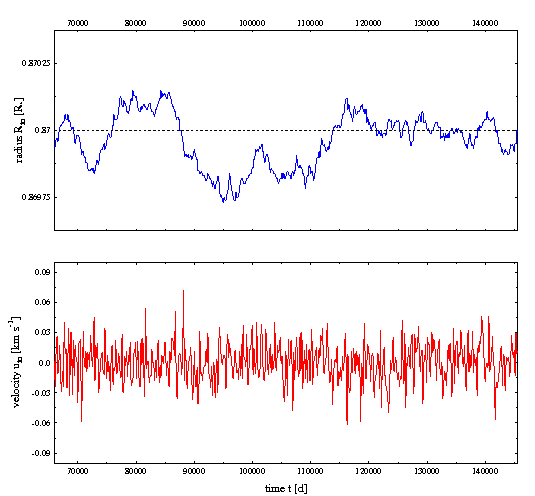}
     \caption{Stochastical movement of the inner boundary $R_{\mathrm{in}}$ about an equilibrium position 
             $R_0=0.87\, R_*$ (upper panel) and associated velocity $u_{\mathrm{in}}$ (lower panel) 
             for the reference model, stochastically excited with strength $\sigma=0.1\,\mathrm{km\,s^{-1}}$.}
    \label{bsp_stochastic}
  \end{figure}

  To examine the eigendynamics of a CDS around a standard or low luminous AGB star (e.g. with the reference 
  stellar parameters: $L_* = 9\times 10^3\,\mathrm{L_{\odot}}$, $M_* = 1\,\mathrm{M_{\odot}}$, $T_* = 2,5\times 
  10^3\,\mathrm{K}$, and $C/O = 1.75$), the additional mechanism of pulsation is adopted to be irregular 
  (stochastically determined), such as is generated by convection in a red giant atmosphere. The reason 
  for this assumption is to avoid the dominance of the stellar pulsation on the CDS dynamics. The shell has 
  been excited with different noise intensities $\sigma=(0.1, 0.5,1.0,2.0)\ \mathrm{km}\,\mathrm{s}^{-1}$ 
  and sequences from up to 24 models were calculated (cf.~Table \ref{sum_stochastic}), each with the 
  same set of fundamental stellar parameters, but for different realisations of the stochastic process as 
  described in Sect.~\ref{stochastic}. 
  The stochastic movement of the inner boundary $R_{\mathrm{in}}(t)$ and the corresponding velocity $u_{\mathrm{in}}(t)$ 
  for an arbitrary stochastic realisation with intensity $\sigma=0.1\,\mathrm{km\,s^{-1}}$ are displayed in Fig. 
  \ref{bsp_stochastic}. 

\begin{figure*}[htb]
   \includegraphics[angle=0,width=0.48\linewidth]{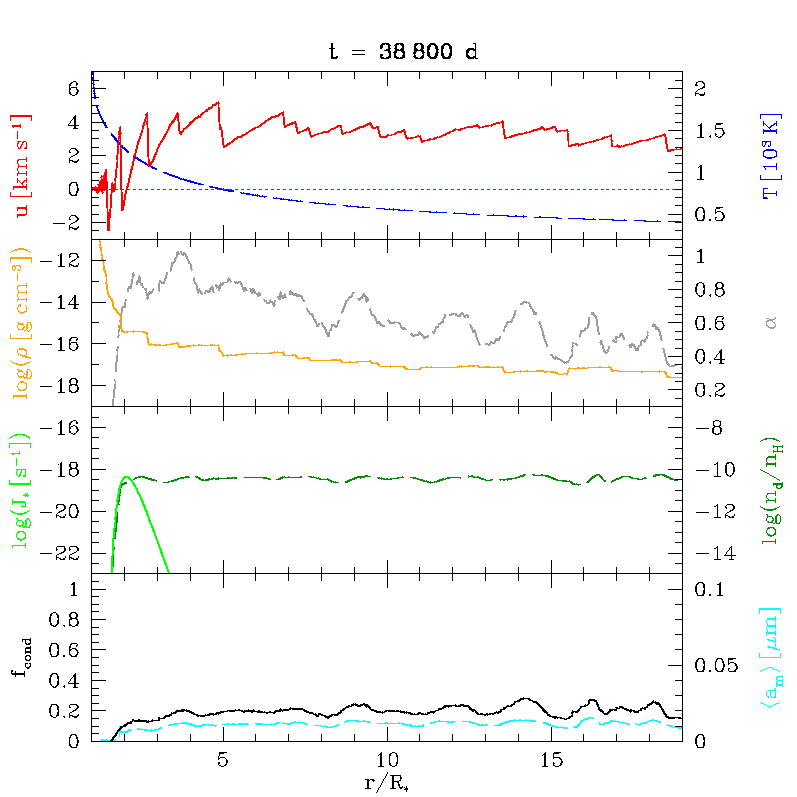}
   \hfill
   \includegraphics[angle=0,width=0.48\linewidth]{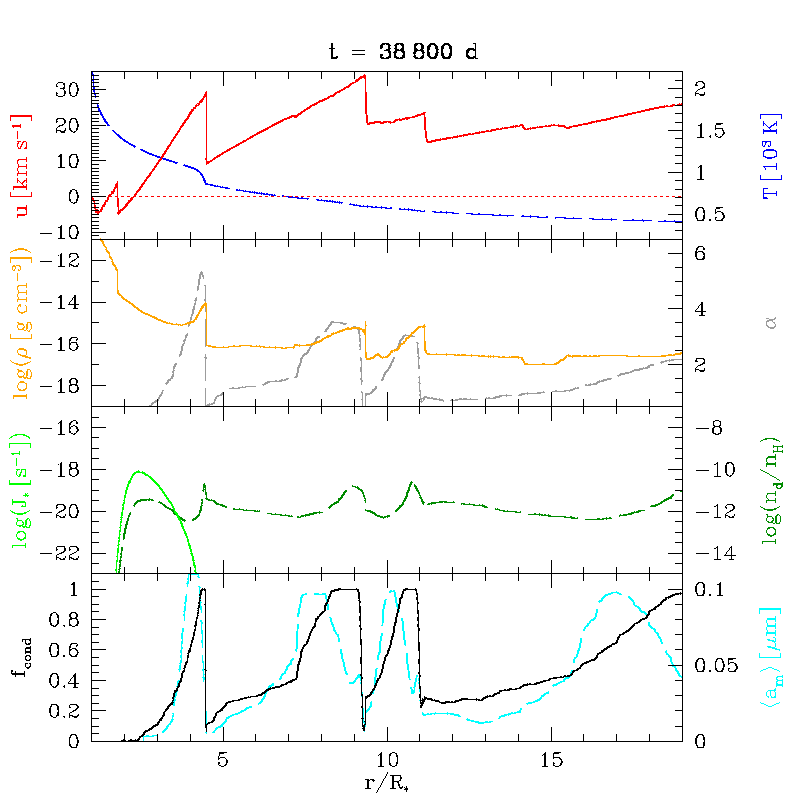}
   \caption{Radial structure of stochastically excited CDSs with excitation strength $\sigma=0.1\,\mathrm{km\,s^{-1}}$ 
           (left) and $\sigma=1.0\,\mathrm{km\,s^{-1}}$ (right).
           \textit{Upper panel}: gas velocity $u$ (solid red line), gas temperature $T$ (dashed blue line). 
           \textit{Second panel}: mass density $\rho$ (solid orange line), radiative acceleration on dust 
           in units of the local gravitational deceleration $\alpha$ (dashed grey line). 
           \textit{Third panel}: stationary nucleation rate $J_*$ (solid light green line), number of dust grains 
           per H-atom $n_{\mathrm{d}}/n_\mathrm{H}$ (dashed dark green line). 
           \textit{Lower panel}: degree of condensation of carbon $f_{\mathrm{cond}}$ (solid black line), 
           mean particle radius $\langle a_{\mathrm{m}}\rangle$ (dashed cyan line).}
   \label{cds_radial_noise}
  \end{figure*}

  Figure \ref{cds_radial_noise} shows snapshots of the radial envelope structure from the stochastically excited model 
  calculation at time $t=38\,800\,\mathrm{d}$ for the intensity strengths $\sigma=0.1\,\mathrm{km\,s^{-1}}$ (left) and 
  $\sigma=1.0\,\mathrm{km\,s^{-1}}$ (right) for various physical quantities describing a CDS. 
  The higher input of mechanical energy in the model with $\sigma=1.0\,\mathrm{km\,s^{-1}}$ can be seen immediately 
  from the higher amplitude of the innermost shock compared to the model with $\sigma=0.1\,\mathrm{km\,s^{-1}}$ 
  (uppermost panels). Since the density scale height is the same for both models in the inner region, the higher 
  energy input yields a smoother transition to the wind region (second panels), i.e. a more levitated atmosphere,  
  hence more favourable conditions for dust formation.
  The outflow is initiated by the pulsation but is accelerated by the dust. The stellar radiation is blocked by the 
  opacity of the instantly formed shell, thereby heating the material behind it, which is located at smaller radii. 
  This backwarming effect prevents new dust from forming and demonstrates that formation and growth of grains also 
  strongly influences the thermodynamical shell structure.
  As depicted in the lowermost left panel for $\sigma=0.1\,\mathrm{km}\,\mathrm{s}^{-1}$, the radial structure is 
  rather smooth with an almost constant degree of condensation $f_{\mathrm{\mathrm{cond}}}\approx0.2$; i.e., only 20\% 
  of the condensable carbon has actually condensed into solid particles.

  This results in an optically thin, slowly dust-driven wind with a time-averaged mass loss rate of $\langle\dot{M}
  \rangle$\footnote{$\langle x\rangle = 1/T \int_{t}^{t+T}x(t')\,dt'$. This nomenclature denotes averaging of quantity 
  at a fixed radial position of $r=25\,\mathrm{R_*}$ over a time interval T of at least one pulsation period $P$.}$
  \approx 9.5\times 10^{-8}\ \mathrm{M_{\odot}}\,\mathrm{yr}^{-1}$ and terminal velocity of $\langle u_{\infty}\rangle
  \approx 3.6\ \mathrm{km\,s^{-1}}$. 
  For $\sigma=1.0\ \mathrm{km}\,\mathrm{s}^{-1}$, the wind becomes rather irregular, but shows the well-known onion-like 
  shell structure (lowermost right panel). The calculated mass loss rate is $\langle\dot{M}\rangle= 4.5\times10^{-6}\ 
  \mathrm{M_{\odot}}\,\mathrm{yr}^{-1}$, and the final outflow velocity shows a typical value of several tens of kilometres 
  per second for an AGB star, namely $\langle u_{\infty}\rangle\approx 25.4\ \mathrm{km\,s^{-1}}$. For both excitation 
  strengths the nucleation zone is localised around $(2-3)\ \mathrm{R_*}$ (third panels).\\

\begin{figure*}[htb]
   \begin{minipage}[t]{\linewidth}
    \includegraphics[width=0.45\linewidth]{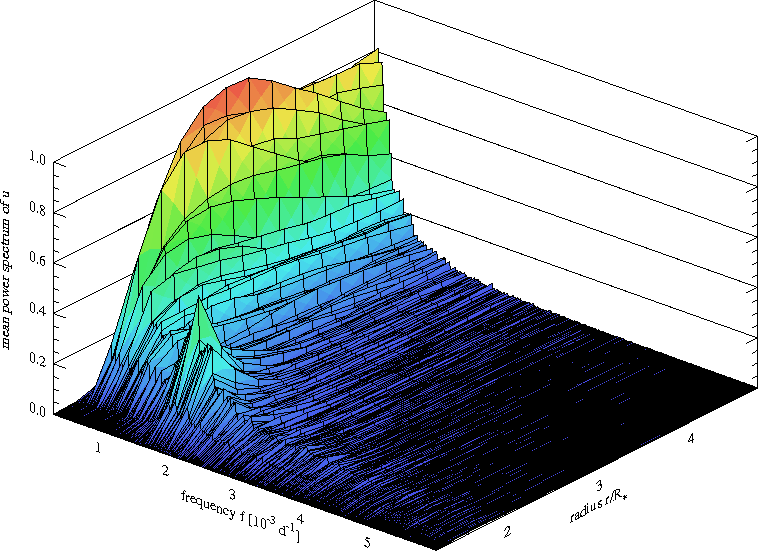}
    \hfill
    \includegraphics[width=0.45\linewidth]{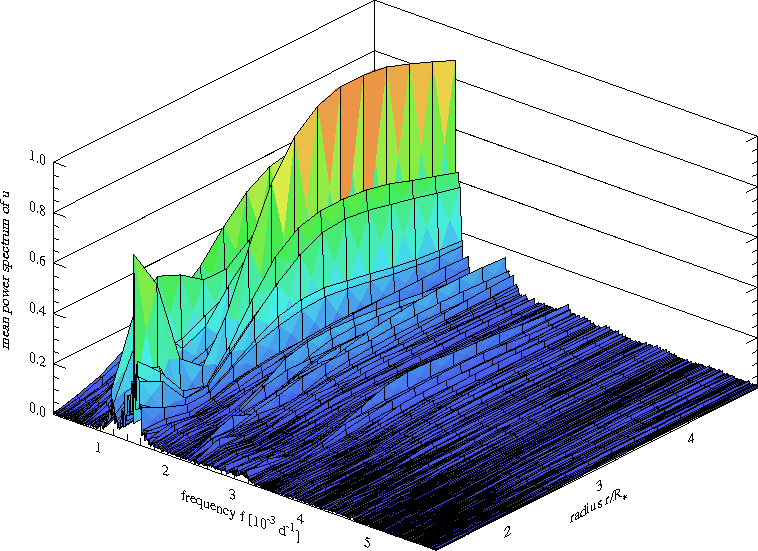}
   \end{minipage}
    \caption{Normalised mean power spectra of the radial gas velocity $u$ caused by stochastically excitation of 
            intensity $\sigma=0.1\,\mathrm{km\,s^{-1}}$ (left) and $\sigma=1.0\,\mathrm{km\,s^{-1}}$ (right).}
    \label{3dnoise}
  \end{figure*}

   These stochastically excited CDSs were subjected to a Fourier transform. Since the gas velocity $u(r,t)$ of a CDS 
   outflow is usually taken to be a tracer of (multi)periodicity, we focused on a detailed analysis of the velocity 
   structure. Regularly sampled with an interval $\Delta t$ over a time domain of $t_{max}=160\,000\,\mathrm{d}$, the 
   time-dependent velocity $u(r,t)$ was transformed at each fixed radial position r for $1\ \mathrm{R_*} \le r\le 5\,
   \mathrm{R_*}$, i.e. within the dust-forming zone. The spatial resolution was chosen to be $\Delta r=0.25\,\mathrm{R_*}$.
   To identify whether spurious frequencies are generated either by interaction between the grid with numbers of
   gridpoints $j_{\mathrm{max}}$, grid remapping $rezoinc$\footnote{For details of grid redistribution see \cite{fleischer94}} 
   or by the choice of the sampling interval $\Delta t$, we studied several numerical parameter combinations.
   These studies suggest the following optimal numerical parameter set for our reference CDS: $N=512$, $j_{\mathrm{max}}=4098$, 
   $rezoinc=0.4\,\mathrm{P}$, and $\Delta t=83\,\mathrm{d}$, which yields a Nyquistfrequency of $f_\mathrm{c}=6\times 10^{-3}\,
   \mathrm{d^{-1}}$.
   Averaging all stochastic realisations provides an overall mean power spectrum of the local expansion gas velocity $u$, 
   which is presented in Fig.~\ref{3dnoise} for the same excitation strengths $\sigma=(0.1, 1.0)\ \mathrm{km}\,\mathrm{s}^{-1}$ 
   as discussed in Fig.~\ref{cds_radial_noise}. See also Table \ref{sum_stochastic}, which lists the model parameters of 
   excitation strength $\sigma$, the number of ensemble members $No.$, and the resulting quantities: ensemble-averaged 
   CDS-eigenmodes $f_{\kappa}$, time-averaged mass loss rate $\langle\dot{M}\rangle$, outflow velocity $\langle u_{\infty}\rangle$, 
   and dust-to-gas mass ratio $\langle\rho_d/\rho_g\rangle$ in the dust nucleation zone at $r\approx2.5\,\mathrm{R_*}$.
   \begin{table*}[htb]
     \caption{List of model calculations and resultant quantities for the stochastically excited reference model.}
     \centering
      \begin{tabular}{cccccccccc}
       \hline\hline
       \noalign{\smallskip}
       $\sigma$ & No. & \multicolumn{5}{c}{$f_{\mathrm{\kappa}}^{\ \mathrm{a}}$} & \multicolumn{1}{c}{$\langle\dot{M}\rangle$} & 
       \multicolumn{1}{c}{$\langle u_{\infty}\rangle$} & \multicolumn{1}{c}{$\langle\rho_{\mathrm{d}}/\rho_{\mathrm{g}}\rangle$}\\
       $(\mathrm{km\,s^{-1}})$ &  & \multicolumn{5}{c}{$(10^{-3}\,\mathrm{d^{-1}})$} & $(\mathrm{M_{\odot}\ yr^{-1}})$ & $(\mathrm{km\ s^{-1}})$ & \\
       \hline
       \noalign{\smallskip}
         $0.1$ & $23$ & $0.06$ & $\mathbf{0.51}$ & -  & $1.58$ & -      & $6.6\times 10^{-7}$ & $ 6.0$ & $1.4\times 10^{-3}$\\
         $0.5$ & $13$ & $0.05$ & -  & $\mathbf{1.04}$ & $1.54$ & $2.13$ & $5.0\times 10^{-6}$ & $23.4$ & $2.8\times 10^{-3}$\\
         $1.0$ & $ 9$ & $0.06$ & -  & $\mathbf{1.13}$ & $1.50$ & $2.13$ & $4.8\times 10^{-6}$ & $25.1$ & $3.0\times 10^{-3}$\\
         $2.0$ & $21$ & $0.06$ & -  & $\mathbf{1.19}$ & $1.51$ & $2.22$ & $5.4\times 10^{-6}$ & $25.2$ & $3.0\times 10^{-3}$\\
       \hline
       \noalign{\smallskip}
      \end{tabular}
      \begin{list}{}{}
        \item[$^{\mathrm{a}}$] Values in boldface indicate the dominant mode
      \end{list}
     \label{sum_stochastic}
  \end{table*}

\begin{figure*}
    \centering
     \includegraphics[angle=0,width=\linewidth]{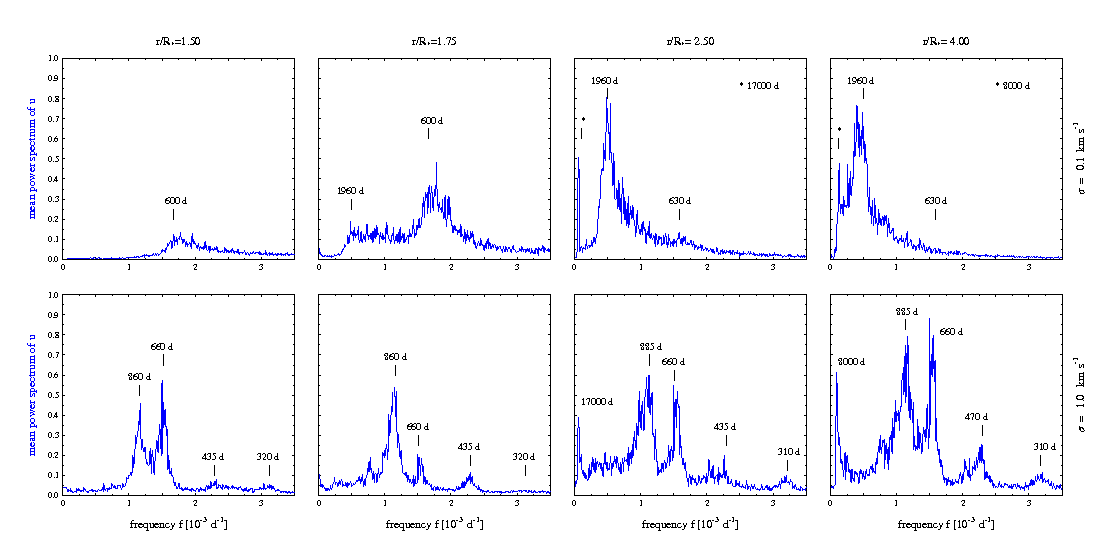}
     \caption{Selected mean power spectrum profiles of the gas velocity $u$ at different radii $r$ 
              for the noise intensities $\sigma=0.1\,\mathrm{km\,s^{-1}}$ 
              (upper panels) and $\sigma =1.0\,\mathrm{km\,s^{-1}}$ (lower panels).}
     \label{noise_schnitt}
  \end{figure*}

  To gain more insight, Fig.~\ref{noise_schnitt} illustrates the power spectra for some selected radial 
  positions within the circumstellar envelope. For small excitation strength $\sigma=0.1\,
  \rm{km\, s^{-1}}$, the response spectra are dominated by three distinct peaks at $f_{\kappa}=
  0.06\times10^{-3}\,\mathrm{d^{-1}}\approx (17\,000\,\mathrm{d})^{-1}$, at $f_{\kappa}=
  0.51\times10^{-3}\,\mathrm{d^{-1}}\approx (1\,960\,\mathrm{d})^{-1}$, and at $f_{\kappa}=
  1.58\times10^{-3}\,\mathrm{d^{-1}}\approx (630\,\mathrm{d})^{-1}$. 
  The amplitude of the third frequency reaches its maximum near the inner edge of the dust nucleation zone at 
  $r/R_{*}=1.75$ and diminishes rapidly with increasing distance from the star. 
  Beyond $r/R_{*}=1.75$, dust formation sets in and the first and second maxima at $f_{\kappa}=0.06\times
  10^{-3}\,\mathrm{d^{-1}}\approx (17\,000\,\mathrm{d})^{-1}$ and at $f_{\kappa}=0.51\times 10^{-3}\,
  \mathrm{d^{-1}}\approx (1\,960\,\mathrm{d})^{-1}$ become visible. 
  They eventually dominate the spectrum for $r/R_{*}>1.75$.

  For the higher excitation strength  $\sigma=1.0\ \rm{km\, s^{-1}}$, the power spectrum looks much richer 
  in detail than for $\sigma=0.1\ \rm{km\,s^{-1}}$ with at least five distinct maxima. Even closer to the 
  star where the medium is too hot to allow for effective dust formation, two strong peaks at $f_{\kappa}=
  1.13\times10^{-3}\,\mathrm{d^{-1}}\approx(885\,\mathrm{d})^{-1}$ and $f_{\kappa}=1.50\times10^{-3}\,
  \mathrm{d^{-1}}\approx(670\,\mathrm{d})^{-1}$ can be seen. Here, the dynamics of the gas are influenced 
  by the backwarming of the dust and the dilution wave originating in the dust nucleation zone (cf. Paper I). 
  With increasing distance from the star, additional maxima appear at $f_{\kappa}= 0.06\times10^{-3}\,
  \mathrm{d^{-1}}\approx(17\,000\,\mathrm{d})^{-1}$, $f_{\kappa}= 2.13\times10^{-3}\,\mathrm{d^{-1}}\approx
 (470\,\mathrm{d})^{-1}$, and $f_{\kappa}=3.2\times10^{-3}\mathrm{d^{-1}}\approx(310\,\mathrm{d})^{-1}$. 

  Similar to the case of a self-excited CDS (Paper I), which can be characterised by an eigenmode $P_{\kappa}$, 
  these frequency maxima are characteristic of the dynamics of envelope around low and standard luminous LPVs 
  and Miras. The result shows that a CDS does not simply mimic an external excitation but has its own eigenmodes,  
  which are determined by the complex interplay between dust formation, growth, radiative transfer, and hydrodynamics. 
  It has also become apparent that the response of the CDS strongly depends on the excitation strength. It is assumed 
  that the maxima at $f_{\kappa}=1.58\times 10^{-3}\,\mathrm{d^{-1}}\approx(630\,\mathrm{d})^{-1}$ for 
  $\sigma=0.1\,\rm{km\,s^{-1}}$ and at $f_{\kappa}=1.50\times10^{-3}\mathrm{d^{-1}}\approx(670\,\mathrm{d})^{-1}$ 
  for $\sigma=1.0\,\rm{km\,s^{-1}}$ belong to the same eigenmode, whose frequency is detuned towards 
  lower values as the strength of the excitation increases. 

\subsection{Dynamics of periodically excited envelopes}
  This section examines the response of the CDS to an external excitation by a sinusoidally oscillating stellar 
  atmosphere (cf.~Sect.~\ref{monoperiodic}). Owing to the inherent non-linearities of a dust-forming system, the 
  CDS does not generally oscillate with the applied frequency but usually exhibits a much more complex behaviour.
  Depending on the excitation frequency, it may exhibit an irregular or periodic behaviour.
  Since the dynamics in CDSs are strongly influenced by the presence of dust, we focus our investigation 
  not only on the dust-formation zone, but also on the physical quantities responsible for providing favourable
  conditions for dust formation, such as gas velocity $u$, density $\rho$, and temperature $T$.\\
\subsubsection{Excitation frequency}\label{a14iresonanz}
\begin{figure}[htb]
    \centering
    \includegraphics[width=9.5cm]{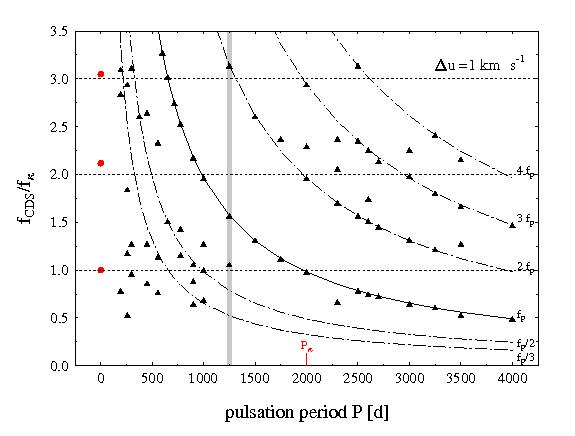}
    \caption{Most dominant frequencies of the CDS $f_{\mathrm{CDS}}$ normalised by the eigenmode $f_{\kappa}=0.51\cdot10^{-3}\,
             \mathrm{d^{-1}}$ (triangles), periodically excited with various stellar pulsation periods $P$ but fixed amplitude 
             $\Delta u = 1.0\,\mathrm{km\,s^{-1}}$, at the dust nucleation zone $r\approx2.5\,\mathrm{ R_*}$. The set of 
             eigenfrequencies are indicated by circles. The eigenperiod is labelled with $P_{\kappa}$. The excitation frequencies 
             lie on the solid line and their harmonics on the dash-dotted lines. The grey line separates the irregular (left) 
             from the pulsation-dominated domain (right).}
     \label{plot_frequency}
  \end{figure}

  Figure \ref{plot_frequency} shows the most prominent shell modes normalised by the eigenfrequency $f_{\mathrm{CDS}}/
  f_{\kappa}$ in the dust nucleation zone at $r\approx2.5\,\mathrm{R_*}$ as a function of the excitation period $P$
  in the case of our reference model. Once again, the perturbation force was taken as the sine of period $P$ according 
  to Eqs.~(\ref{pulsation_r}) and (\ref{pulsation_v}), whereas the amplitude was held at $\Delta u=1.0\,\mathrm{km}\,
  \mathrm{s}^{-1}$.

  The excitation frequency $f_{\mathrm{p}}=1/P$ itself and some of its harmonics are recognisable for all periods $P$, 
  whereas the determined eigenmodes $f_{\kappa}$ are only visible for small excitation periods $P\lesssim 1\,250\,\mathrm{d}$.
  \begin{itemize}
   \item \textbf{Irregular domain}\\
      For small and intermediate excitation periods $P\lesssim 1\,250\,\rm{d}$, the power spectra show a rather 
      extended continuum with a number of distinct maxima. Some of these could be identified as eigenmodes of the 
      CDS or as their harmonics. Also, the subharmonics $f_{\mathrm{p}}/2$ and $f_{\mathrm{p}}/3$ of the excitation 
      frequency $f_{\mathrm{p}}=1/P$ show up.
      As the pulsation period $P$ increases, the excitation period and its harmonics become clearly
      recognisable. In contrast, the amplitude of the envelope's eigenmode mostly decreases as $P$ increases. 
      Within this domain a kind of resonance phenomenon between excitation and envelope period occurs. 
      The envelope reacts multiperiodically for a sensitive range of excitation period P and strength $\Delta u$.\\
   \item \textbf{Pulsation-dominated domain}\\
     For longer periods $P\gtrsim 1\,250\,\rm{d}$, the CDS becomes finally enslaved by the excitation force.
     The most prominent feature of the spectrum is the pulsation mode supplemented by a set of its harmonics. 
  \end{itemize}

  In general, the final velocity $\langle u_{\infty}\rangle$ has a fairly high value with a maximum at 
  $\langle u_{\infty}\rangle=32\,\mathrm{km\, s^{-1}}$ for $P=600\,\mathrm{d}$. With increasing period $P$, the velocity 
  continuously slopes downward to $\langle u_{\infty}\rangle= 19\,\mathrm{km\,s^{-1}}$ at period $P=4\,500\,\mathrm{d}$. 
  The corresponding mass loss rates show the same trend. The maximum mass loss rate is $\langle\dot{M}\rangle=1.8\times10^{-5}\,
  \mathrm{M_{\odot}\, yr^{-1}}$ for $P=600\,\mathrm{d}$, and the minimum $\langle\dot{M}\rangle=2.6\times 10^{-6}\,\mathrm{M_{\odot}\,
  yr^{-1}}$ for $P=4\,500\,\mathrm{d}$ (cf. Table \ref{table_frequency}).
  \begin{table}[htb]
    \caption{List of resultant quantities from a CDS periodically disturbed.} 
    \centering
     {\small
      \begin{tabular}{p{0.8cm}rcp{0.75cm}ccc}
       \hline\hline
       \noalign{\smallskip}
        \multicolumn{1}{c}{$P$} & \multicolumn{3}{c}{$\mathrm{f_{CDS}^{\ a}}$} & $\langle\dot{M}\rangle$ & $\langle u_{\infty}\rangle$ & 
        $\langle\rho_{\mathrm{d}}/\rho_{\mathrm{g}}\rangle$\\
        \multicolumn{1}{c}{$(\mathrm{d})$} &  \multicolumn{3}{c}{$(10^{-3}\,\mathrm{d^{-1})}$} & $(\mathrm{M_{\odot}\ yr^{-1}})$ & $(\mathrm{km\ s^{-1}})$ &\\
        \hline
        \noalign{\smallskip}
        $\ \ 194$ & $0.40$ & $1.45$ & $\mathbf{1.58}$             & $9.5\times 10^{-6}$ &  $27.5$  &   $3.1\times 10^{-3}$\\
        $\ \ 259$ & $0.27$ & $\mathbf{0.60}$ & $0.94$             & $1.1\times 10^{-5}$ &  $26.4$  &   $3.8\times 10^{-3}$\\
        $\ \ 300$ & $\mathbf{0.49}$ & $0.65$ & $1.59$             & $9.4\times 10^{-6}$ &  $25.6$  &   $3.5\times 10^{-3}$\\
        $\ \ 376$ & $\mathbf{1.33}$ & $\underline{2.66}$ & $3.99$ & $1.0\times 10^{-5}$ &  $26.4$  &   $4.1\times 10^{-3}$\\
        $\ \ 450$ & $\mathbf{0.44}$ & $0.65$ & $1.35$             & $1.1\times 10^{-5}$ &  $25.8$  &   $3.9\times 10^{-3}$\\
        $\ \ 555$ & $0.39$ & $\mathbf{0.58}$ & $1.19$             & $1.6\times 10^{-5}$ &  $26.2$  &   $4.1\times 10^{-3}$\\
        $\ \ 600$ & $\underline{\mathbf{1.67}}$ & $3.33$ & $5.00$ & $1.8\times 10^{-5}$ &  $31.8$  &   $4.2\times 10^{-3}$\\
        $\ \ 650$ & $0.77$ & $\underline{\mathbf{1.54}}$ & $3.08$ & $7.9\times 10^{-6}$ &  $28.9$  &   $4.2\times 10^{-3}$\\
        $\ \ 716$ & $\underline{\mathbf{1.40}}$ & $2.79$ & $4.19$ & $7.5\times 10^{-6}$ &  $26.8$  &   $4.2\times 10^{-3}$\\
        $\ \ 900$ & $0.45$ & $0.54$ & $\underline{\mathbf{1.11}}$ & $9.7\times 10^{-6}$ &  $25.2$  &   $3.2\times 10^{-3}$\\
        $1000$    & $0.35$ & $0.51$ & $\underline{\mathbf{1.00}}$ & $1.7\times 10^{-5}$ &  $24.4$  &   $2.7\times 10^{-3}$\\
        $1250$    & $0.54$ & $\underline{\mathbf{0.80}}$ & $1.60$ & $1.3\times 10^{-5}$ &  $26.3$  &   $4.2\times 10^{-3}$\\[0.2em]
        \hline\\[-0.6em]
        $1500$    & $\underline{\mathbf{0.67}}$ & $1.33$ & $2.00$ & $5.0\times 10^{-6}$ &  $26.8$  &   $3.7\times 10^{-3}$\\
        $1750$    & $\underline{0.57}$ & $\mathbf{1.21}$ & -      & $4.3\times 10^{-6}$ &  $25.7$  &   $4.1\times 10^{-3}$\\
        $2000$    & $\underline{\mathbf{0.50}}$ & $1.00$ & $1.17$ & $3.8\times 10^{-6}$ &  $23.5$  &   $2.3\times 10^{-3}$\\
        $2300$    & $0.34$ & $\mathbf{0.87}$ & $1.05$             & $2.1\times 10^{-6}$ &  $23.1$  &   $2.6\times 10^{-3}$\\
        $2500$    & $\underline{\mathbf{0.40}}$ & $0.80$ & $1.20$ & $1.8\times 10^{-6}$ &  $18.8$  &   $1.4\times 10^{-3}$\\
        $2600$    & $\underline{0.38}$ & $\mathbf{0.77}$ & $0.89$ & $2.2\times 10^{-6}$ &  $18.4$  &   $1.9\times 10^{-3}$\\
        $2700$    & $\underline{\mathbf{0.37}}$ & $0.74$ & $1.09$ & $3.1\times 10^{-6}$ &  $20.4$  &   $2.9\times 10^{-3}$\\
        $3000$    & $\underline{\mathbf{0.33}}$ & $0.67$ & $1.01$ & $4.5\times 10^{-6}$ &  $20.4$  &   $2.7\times 10^{-3}$\\
        $3250$    & $\underline{\mathbf{0.31}}$ & $0.62$ & $0.92$ & $3.6\times 10^{-6}$ &  $16.8$  &   $2.3\times 10^{-3}$\\
        $3500$    & $\underline{0.27}$ & $0.65$ & $\mathbf{0.85}$ & $4.8\times 10^{-6}$ &  $18.1$  &   $1.9\times 10^{-3}$\\
        $4000$    & $\underline{0.25}$ & -      & $\mathbf{0.75}$ & $2.6\times 10^{-6}$ &  $18.8$  &   $2.9\times 10^{-3}$\\
        \hline
        \noalign{\smallskip}
      \end{tabular}
      \begin{list}{}{}
        \item[$^{\mathrm{a}}$] Values in boldface indicate the most dominant mode and underlined values the excitation frequency
      \end{list}
      }
      \label{table_frequency}
  \end{table}

 \begin{figure*}[htbp]
   \centering
   \begin{minipage}[t]{\linewidth}
     \includegraphics[width=0.45\linewidth]{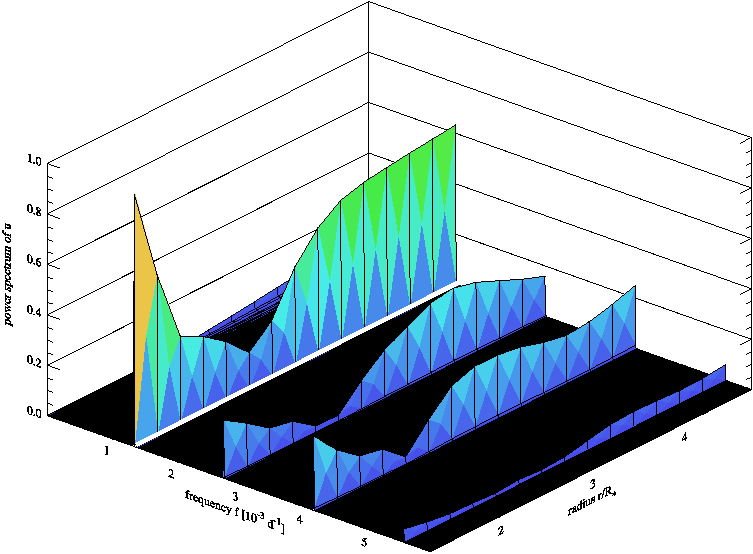}
     \hfill
     \includegraphics[width=0.45\linewidth]{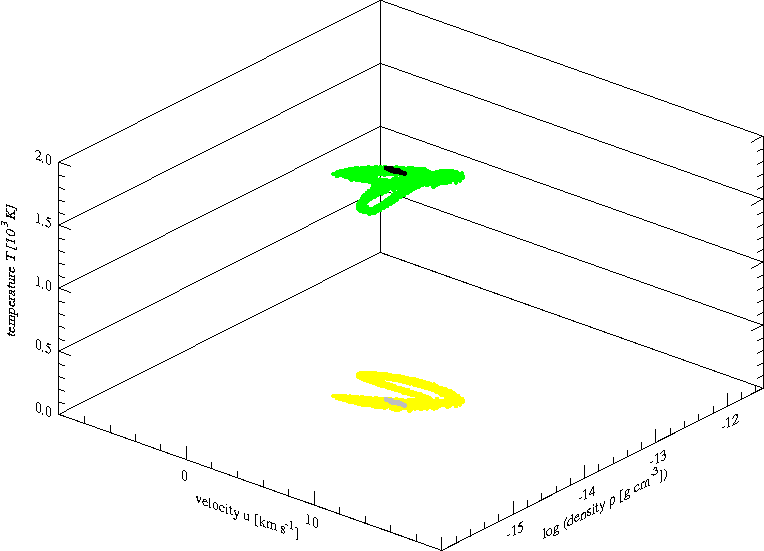}
   \end{minipage}
   \vspace{.5cm}{
   \begin{minipage}[t]{\linewidth}
     \includegraphics[width=.45\linewidth]{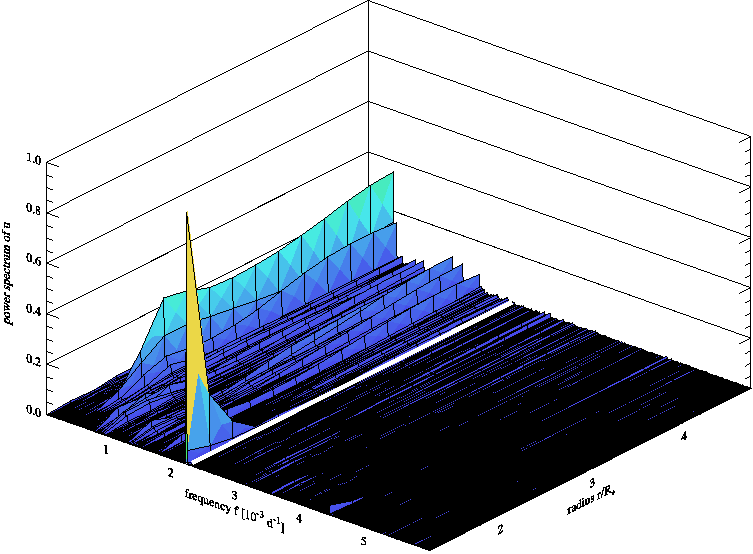}
     \hfill
     \includegraphics[width=.45\linewidth]{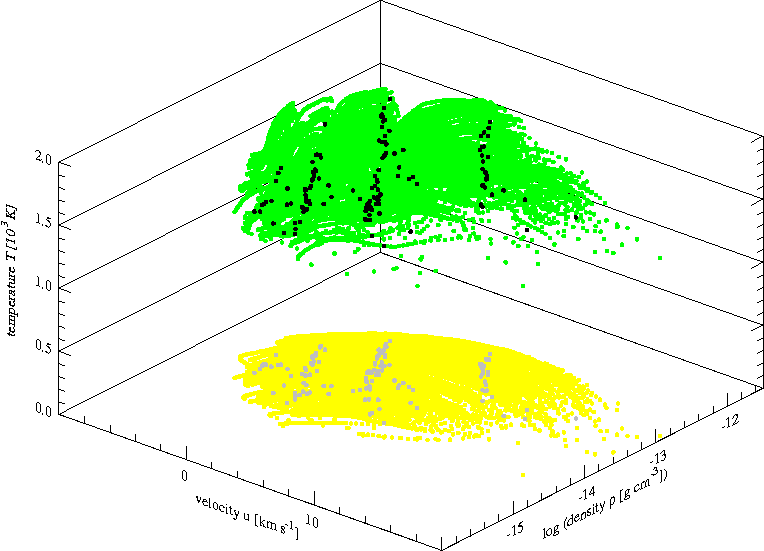}
   \end{minipage}}
   \vspace{.5cm}{
   \begin{minipage}[t]{\linewidth}
     \includegraphics[width=.45\linewidth]{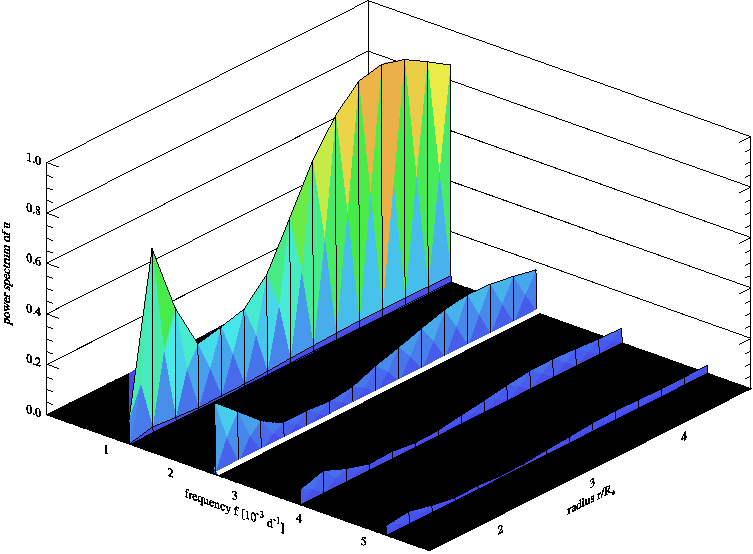} 
     \hfill 
     \includegraphics[width=.45\linewidth]{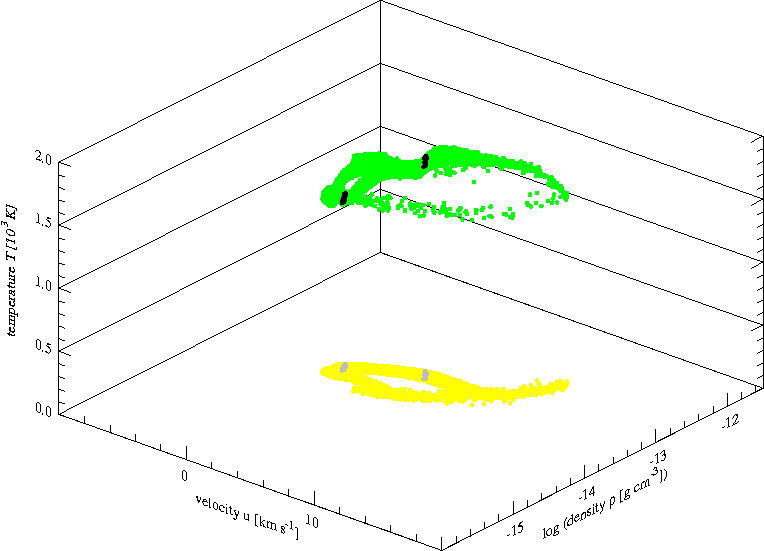}
   \end{minipage}}
   \caption{Power spectra of the local expansion velocity $u$ (left panels) and corresponding maps of 
            the ($u$,$\rho$,$T$) phase space (green) and projection onto ($u$,$\rho$)-plane (yellow) 
            (right panels) in the dust nucleation zone for different monoperiodic excitation 
            periods $P=716\,\mathrm{d}$ (upper), $450\,\mathrm{d}$ (middle), and $376\,\mathrm{d}$ (lower), 
            respectively. The stroboscopic maps were obtained by sampling $[u(n0.02P), \rho(n0.02P), T(n0.02P)]$
            (green) and the Poincar\'e maps by $[u(nP), \rho(nP), T(nP)]$ (black) for $1\le n \le \lfloor 
            t_{max}/P\rfloor, n\in\mathbb{N}$. The excitation frequencies are highlighted in the spectra.}
   \label{phase}
 \end{figure*}

    Figure \ref{phase} demonstrates the basic behaviour of the envelope for some selected monoperiodic excitation 
    periods $P=376\,\mathrm{d}$, $450\,\mathrm{d}$, and $716\,\mathrm{d}$ chosen from the irregular shell dynamics 
    regime. The left panels display the power spectra of the local expansion velocity $u$ by scaling 
    to the overall strongest mode. 
    Finally, the right panels present the stroboscopic maps in the ($u$, $\rho$, $T$)-phase space, supplemented by 
    a projection onto the ($u$, $\rho$)-plane. Both were sampled at  an interval $\Delta t=0.02\,\mathrm{P}$. 
    They also show the corresponding state of the system at a constant phase angle $\phi$, i.e. $\Delta t=P$ 
    (Poincar\'e map) in the dust nucleation zone at $r\approx2.5\,\mathrm{R_*}$.

    As can be seen in the uppermost and lowermost spectra, the CDS reacts with the same dominant mode $f=1.34\times10^{-3}
    \,\mathrm{d^{-1}}=(716\,\mathrm{d})^{-1}$, even though it is excited with different periods $P=716\,\mathrm{d}$ and 
    $P=376\,\mathrm{d}$. This is similar to the resonance phenomenon ($iP=jP_{\mathrm{CDS}}$) studied in paper I 
    for high-luminous Miras and LPVs. However, for the case of standard luminous LPVs and Miras studied in this paper 
    we always found $j=1$, which means that the period of the CDS $P_{\mathrm{CDS}}$ appears as an integer multiple of 
    the excitation period $P$ ($iP=P_{\mathrm{CDS}}$). In the context of circumstellar envelopes, this phenomenon is 
    often referred to as multiperiodicity \citep[cf.][]{winters94,fleischer95,hoefner96}.
    As indicated by the number of the black clusters in the stroboscopic maps, the system reacts monoperiodically 
    (single black cluster, upper right panel, i.e. $i=1$) for an excitation with $P=716\,\mathrm{d}$, whereas the system's 
    response is double-periodic (two black features in the lowermost plot on right, i.e. $i=2$) for an excitation with 
    $P=376\,\mathrm{d}$. In other words, the system returns to the same state after a time interval $\Delta t=iP$.
    The corresponding curves in the $(u,\rho, T)$-hyper plane are restricted and closed. Nevertheless, some parts of 
    the trajectory are passed through extremely rapidly, so that these parts are only depicted by a few points on the 
    corresponding stroboscopic map.

    For $P=450\,\mathrm{d}$, as can be seen from the power spectra in the middle-left panel, the excitation period does 
    not dominate the envelope. The amplitude of the piston frequency $f_{\mathrm{p}}=2.22\times10^{-3}\,\mathrm{d^{-1}}$ 
    is weak, and the frequency spectrum shows up with distinct maxima at $f=0.44\times10^{-3}\,\mathrm{d^{-1}}\approx(2300\,
    \mathrm{d})^{-1}$, which is a characteristical timescale of the CDS and $f=1.35\times 10^{-3}\,\mathrm{d^{-1}}
    \approx (740\,\mathrm{d})^{-1}$, which seems to correspond to the somewhat-shifted feature $P=716\,\mathrm{d}$ providing 
    the resonance case for this model. This is also consistent with the stroboscopic map given in the middle-right panel. 
    In contrast to $P=716\, \mathrm{d}$ and $376\ \mathrm{d}$, the system is not limited to a closed trajectory in the 
    $(u,\rho, T)$-space but rather completely fills out a continuous part of the phase space. However, the dynamics 
    of the envelope once more seem to synchronise with the excitation period to some extent. For a constant phase angle $\phi$ of the 
    stellar pulsation, i.e. $\Delta t=P$, the system stays inside well-defined stripes that cut through 
    the entire phase-space area filled out by the system.\\
  \subsubsection{Excitation strength}
    This section studies the above-mentioned reference model with regard to the response of the CDS for a sinusoidally 
    stellar pulsation with period $P=376\,\mathrm{d}$, i.e. a typical oscillation period of LPVs and Miras, over a 
    sequence of varying strengths, i.e. amplitudes $\Delta u$. The amplitude starts with a value of $\Delta u=1.0\,
    \mathrm{km\,s^{-1}}$, so the disturbance is subsonic, and is then increased up to $\Delta u=9.0\,\mathrm{km\,s^{-1}}$ 
    i.e. into the supersonice regime. Table \ref{strength} lists the results of these calculations. 

    The excitation frequency $f_{\mathrm{p}}=2.66\times10^{-3}\,\mathrm{d^{-1}}=(376\,\mathrm{d})^{-1}$ itself is only 
    noticeable for a low excitation strength of $\Delta u=1.0\,\mathrm{km\,s^{-1}}$. Because weak in power, it vanishes 
    with increasing excitation strengths. The inertial envelope is not able to follow the short excitation period; in 
    fact, this mode at no time does it dominate the envelope dynamics. Actually, the first subharmonic of the excitation 
    frequency $f_{\mathrm{p}}/2=1.33\times10^{-3}\,\mathrm{d^{-1}}\approx(752\,\mathrm{d})^{-1}$ dominates the dynamics. 
    It can be regarded as a detuned envelope-eigenmode $f_{\kappa}= 1.50\times10^{-3}\,\mathrm{d}\approx(670\,\mathrm{d})^{-1}$. 
    Subsequently, the eigenmode at $f_{\kappa}=0.50\times 10^{-3}\,\mathrm{d^{-1}}\approx (2000\,\mathrm{d})^{-1}$ takes 
    over the dominance of the shell dynamics. In general, the CDS-eigenmodes are detuned towards lower frequencies as 
    already shown in the case of a CDS around a high-luminous LPV (cf. paper I).  
    The stellar pulsation leads to a levitation of the atmosphere; i.e., it lifts the material out of the gravitational 
    field into the dust-forming zone. With increasing strength of the pulsation amplitude $\Delta u$, the additional 
    input of energy and momentum leads to an enhanced density in the dust-forming region and supports further dust 
    formation rather than accelerating the wind. Consequently, the stellar wind becomes far more massive, and the mass loss 
    rate increases from $\langle\dot{M}\rangle=1.0\times10^{-5}\,\mathrm{M}_{\odot}\,\mathrm{yr}^{-1}$ for $\Delta u=1\,
    \mathrm{km\,s^{-1}}$ to $\langle\dot{M}\rangle=7.4\times10^{-5}\,\mathrm{M}_{\odot}\,\mathrm{yr}^{-1}$ for $\Delta 
    u=9\,\mathrm{km\,s^{-1}}$ at nearly constant velocity. As the wind becomes more intense, the periodic depletion in 
    dust-forming material also increases, and it takes more time to re-enrich the nucleation zone. The result is a longer 
    shell period as seen in Table \ref{strength}. 

    \begin{table}[htb]
     \caption{Parameter study for various excitation strength $\Delta u$ with fixed period $P=376\,\mathrm{d}$.}
      \centering
       {\small
       \begin{tabular}{crcp{0.75cm}ccc}
        \hline\hline
        \noalign{\smallskip}
        \multicolumn{1}{c}{$\Delta u$} & \multicolumn{3}{c}{$f_{\mathrm{CDS}}^{\ \mathrm{a}}$} & $\langle\dot{M}\rangle$ & 
        $\langle u_{\infty}\rangle$ & $\langle\rho_{\mathrm{d}}/\rho_{\mathrm{g}}\rangle$\\
        \multicolumn{1}{c}{$(\mathrm{km\,s^{-1}})$} & \multicolumn{3}{c}{$(10^{-3}\,\mathrm{d^{-1})}$} & $(\mathrm{M_{\odot}\ yr^{-1}})$ & $(\mathrm{km\ s^{-1}})$ &\\ 
        \hline
        \noalign{\smallskip}
         $1.0$ & $\mathbf{\uuline{1.33}}$ & $\underline{2.66}$ & $3.99$  &  $1.0\times 10^{-5}$ & $26.4$ & $4.1\times 10^{-3}$\\
         $2.0$ & $\mathbf{0.53}$ & $1.06$ & $\uuline{1.33}$              &  $3.8\times 10^{-5}$ & $29.9$ & $4.3\times 10^{-3}$\\
         $3.0$ & $\mathbf{0.46}$ & $0.88$ & $\uuline{1.33}$              &  $2.5\times 10^{-5}$ & $25.9$ & $3.3\times 10^{-3}$\\
         $4.0$ & $\mathbf{0.47}$ & $0.90$ & $\uuline{1.35}$              &  $8.5\times 10^{-6}$ & $25.6$ & $3.2\times 10^{-3}$\\
         $5.0$ & $0.47$          & $0.93$ & $\mathbf{\uuline{1.33}}$     &  $2.1\times 10^{-5}$ & $28.2$ & $3.7\times 10^{-3}$\\
         $6.0$ & $\mathbf{0.46}$ & $0.79$ & $\uuline{1.30}$              &  $4.1\times 10^{-5}$ & $30.0$ & $4.0\times 10^{-3}$\\
         $7.0$ & $\mathbf{0.43}$ & $0.93$ & $\uuline{1.29}$              &  $1.9\times 10^{-5}$ & $28.4$ & $3.7\times 10^{-3}$\\
         $8.0$ & $\mathbf{0.45}$ & $0.86$ & $\uuline{1.33}$              &  $1.2\times 10^{-5}$ & $28.4$ & $3.5\times 10^{-3}$\\
         $9.0$ & $\mathbf{0.44}$ & $0.87$ & $1.69$                       &  $7.4\times 10^{-5}$ & $30.6$ & $3.5\times 10^{-3}$\\
        \hline
        \noalign{\smallskip}
       \end{tabular}
       \begin{list}{}{}
         \item[$^{\mathrm{a}}$]\footnotesize{Values in boldface indicate the most dominant mode, whereas single and double underlined values 
              refer to the excitation frequency and its first harmonic, respectively.}
       \end{list}
       \label{strength}
      }
    \end{table}

    \begin{figure*}[htb]
     \centering
     \begin{minipage}[t]{\linewidth}
      \includegraphics[width=0.45\linewidth]{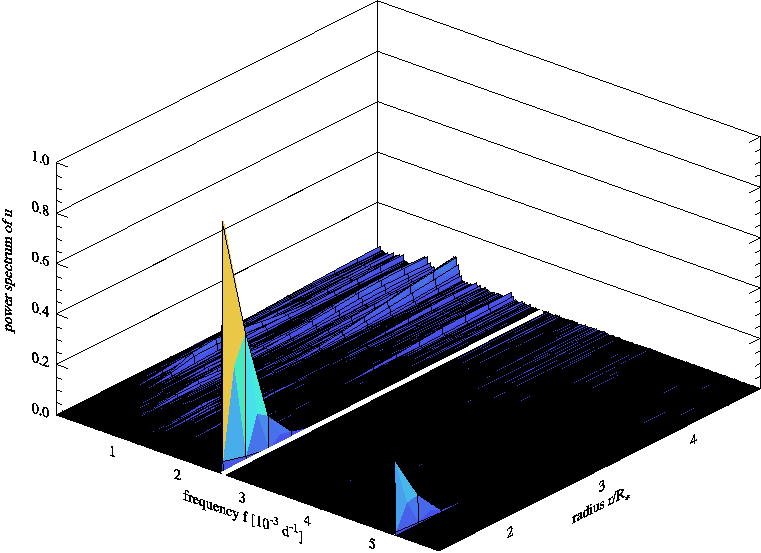}
      \hfill
      \includegraphics[width=0.45\linewidth]{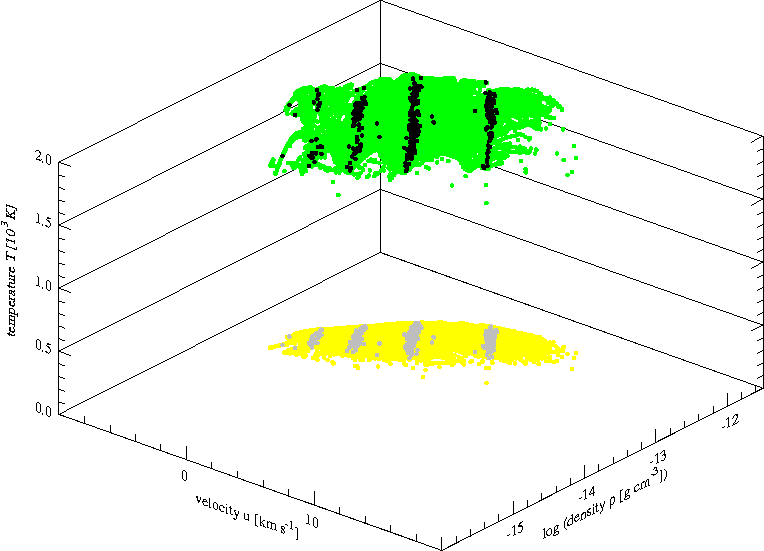}
     \end{minipage}
     \caption{Power spectrum and stroboscopic maps for a CDS excited with a period $P=376\,\mathrm{d}$ and strength 
              $\Delta u = 9\,\mathrm{km\, s^{-1}}$. The plot parameters for the maps are the same as in Fig.~\ref{phase}.}
     \label{amplitudep376}
    \end{figure*}

    Figure \ref{amplitudep376} depicts the spectrum and maps of the reference CDS excited with period $P=376\,\mathrm{d}$ 
    and amplitude $\Delta u = 9\,\mathrm{km\,s^{-1}}$. With increasing amplitude the CDS behaviour changes from multiperiodic 
    to irregular, as can be seen in the stroboscopic maps (cf. Fig.~\ref{phase}, lowermost panels, right and Fig.~\ref{amplitudep376}, 
    right). The double periodicity breaks up. As can be seen in the power spectrum, the timescale of the stellar pulsation 
    (excitation modes) imprints their dynamics onto the envelope close to the star. With onset of the dust formation at $r\approx 
    2.5\,\mathrm{R_*}$ a continuous power spectrum appears rather than distinct modes in terms of the radial excitation frequency.
%
%
%
\section{Summary and outlook}\label{summary}
In this article we have investigated the dynamics of envelopes around standard and low-luminous, carbon-rich 
AGB stars. We analysed the numerical solutions of the coupled equation system of hydrodynamics, 
thermodynamics, dust nucleation and growth, and radiative transfer by means of power spectra and 
stroboscopic maps of quantities, such as radial outflow velocity, temperature, and density. 

As shown for the case of CDSs around highly luminous AGBs, the presence of dust influences the dynamical and 
energetic behaviour of the entire system. Firstly, it amplifies the momentum coupling 
between matter and radiation field, and secondly, it enforces an additional dynamical behaviour (exterior 
$\kappa$-mechanism) besides the pulsation to the dust envelope. The consequence of the coupling is the development 
of an envelope dynamics, which is in fact excited by the momentum and energy input by the stellar pulsation, but is 
operating on the timescales of the physical processes that determining the envelope. These dynamics can be characterised 
by a set of autonomous frequencies that are not identical to the frequency of the pulsation.

For CDSs around low and standard luminous AGBs, such distinct frequencies are not obvious. 
To study the shell eigendynamics, the system has to be supplied with energy and momentum. This we have 
done by introducing an additional force, regarded stochastic, e.g. provided by convection 
in the giant's atmosphere. The goal is to supply enough energy to introduce a CDS generating mass loss, 
but also to minimise the influence on the CDS's eigendynamics.

When applying this method with different stochastic noise intensities to a reference CDS, the power spectra of 
the radial velocity show up as a set of eigenmodes of about $10^3\,\mathrm{d}$, which represents 
the dynamical timescale of dust formation and element enhancement in the dust formation zone.
To study the interaction between the envelopes's eigendynamics and stellar pulsation, a parameter 
study of various stellar excitation periods and strengths was carried out. Depending on the stimulation, 
the system reacts periodically, multiperiodically, or irregularly.
For short excitation periods, the considered power spectra suggest that the CDS models tend to 
be semi-regular or even chaotic. The timescale of the excitation is much smaller than the timescale of 
the various physical processes (in particular the timescale of dust formation) in the envelope, so that 
the shell is not able to follow the excitation. 
For long periods the system is dominated by the external excitation. In contrast to envelopes around high 
luminous AGBs, no eigenmode-dominated domain could be found. In a closer examination of excitation periods 
from the irregular domain, we found the same resonance phenomenon ($iP=jP_{CDS}$) as in the case of 
envelopes around high-luminous LPVs and Miras. However, it turns out that the integer j always equal 1 
for standard luminous AGBs. 

Combined with the analysis method of shells around highly luminous AGBs already presented in a previous article, 
we are now able to start a systematic stellar parameter study. In order to compare the results with observations, this planned 
study will also analyse synthetic lightcurves and spectra.
%
%
%
%
%
%

\end{document}